\documentclass{aa}  
\usepackage{graphicx}
\usepackage{natbib}
\usepackage{txfonts}
\usepackage[]{hyperref}
\renewcommand{\eqref}[1]{Eq.~(\ref{#1})}
\newcommand{\figref}[1]{Fig.~\ref{#1}}
\newcommand{\secref}[1]{Sect.~\ref{#1}}
\newcommand{\tabref}[1]{Tab.~\ref{#1}}
\usepackage[dvipsnames]{xcolor}

\defcitealias{neumann1950shocks}{VNR50}
\defcitealias{tscharnuter1979artificial_viscosity}{TW79}
\defcitealias{stone1992zeus2d}{SN92}

\begin{document}

\title{Hydrodynamical simulations with strong indirect terms in \textsc{Fargo}-like codes}
\subtitle{Numerical aspects of non-inertial frame and artificial viscosity}

\author{Lucas M. Jordan
	\inst{1}
	\and
	Thomas Rometsch
	\inst{2}
}

\institute{Institut für Astronomie und Astrophysik, Universit\"at T\"ubingen,
	Auf der Morgenstelle 10, 72076 T\"ubingen, Germany
	\and
	Zentrum für Astronomie (ZAH), Institut für Theoretische Astrophysik (ITA), Universit\"at Heidelberg,
	Albert-Ueberle-Str. 2, 69120 Heidelberg, Germany\\
	\email{lucas.jordan@uni-tuebingen.de}
}

\date{Received \today}


\abstract
{
	Binary star systems allow us to study the planet formation process under extreme conditions.
	In the early stages, these systems contain a circumbinary disk and a disk around 
	each star. To model the interactions between these disks in the frame of one of the stars,
	strong fictitious forces must be included in the simulations.
	The original \textsc{Fargo} and the \textsc{Fargo3D} codes fail to correctly simulate such
	systems if the indirect term becomes too strong.
}
{
	We present a different way to compute the indirect term 
	which, together with a tensor artificial viscosity prescription,
	allows the \textsc{Fargo} code to simulate the
	circumbinary disks in a non-inertial frame of reference.
	In this way, the \textsc{Fargo} code can be used to study
	interactions between circumstellar and circumbinary disks.
}
{
	We first evaluate the accuracy of the standard implementation and our proposed 
	indirect term prescription using a simple N-body test case. We then
	analytically evaluate the effect of the default artificial viscosity 
	used in the \textsc{Fargo} code in the limit of large distances to the N-body system.
	Finally, we evaluate the effects of the different prescriptions 
	by performing hydrodynamical simulations in a non-inertial frame 
	of reference.
}
{
	By updating the indirect term prescription and the artificial viscosity,
	we were able to successfully simulate a circumbinary disk in a frame that is centered on 
	the less massive star.
	We find that updating the indirect term becomes relevant when the indirect term becomes stronger than the direct 
	gravitational forces, which occurs for mass ratios of $q \gtrsim 5\%$.
	The default artificial viscosity used in the \textsc{Fargo} code
	inherently produces artificial pressure in a non-inertial frame of reference
	even in the absence of shocks. This leads to artificial mass ejection from the 
	Hill sphere, starting at brown dwarf masses ($q \gtrsim 1\%$).
	These problems can be mitigated by using a tensor artificial viscosity formulation.
	For high mass ratios, $q \gtrsim 1\%$, it is also becomes important to initialize the disk 
	in the center-of-mass frame.

	We expect our proposed changes to be relevant for other grid-based hydrodynamic codes
	where strong indirect terms occur, or for codes that use artificial viscosity.
}
{}


\maketitle
%
%
\section{Introduction}
When studying astrophysical hydrodynamics simulations, it is of interest to
conduct them in a non-inertial frame of reference, e.g.\, to center the simulation
on one star of a binary system to study its circumstellar disk. The non-inertial frame of
reference introduces fictitious forces that must be taken into account in the simulation.

The indirect term is supposed to compensate for the forces
acting on the center of the frame of reference so that it does not move.
In extreme cases, the force due to the indirect term dominates the total force acting on the gas.
Under such conditions, an overly simplistic approach for handling the indirect
term can become the limiting factor in the accuracy of the entire simulation.

In this paper, we discuss the artificial viscosity and
the indirect term used in the \textsc{Fargo} and \textsc{Fargo3D} codes
\citep{masset2000fargo,benitez2016fargo3d} and show
that they cause problems when simulating extreme setups.
We then propose a more accurate method for handling the indirect term and
the use of the tensor artificial viscosity by \citet{tscharnuter1979artificial_viscosity}
and show that they are more accurate at simulating systems with strong indirect terms.
The methods presented in this paper have already been mentioned in \citet{rometsch2024fargocpt}.
In this paper, we explain the methods in detail and explore when they become necessary.

In \secref{sec:nbody} we compare two strategies for calculating the
indirect term in a binary star system by testing how well the primary component
of the binary remains fixed in a reference frame centered on the primary.
We discuss two
forms of artificial viscosity that are
commonly used in grid codes in \secref{sec:art_vis},
and then use the proposed changes to run hydrodynamical simulations
with strong indirect terms in \secref{sec:binary}.
We then evaluate the relevance of our modifications to planetary disk simulations
in \secref{sec:planet} and discuss our results in \secref{sec:summary}.
\section{N-body example on applying the indirect term}
\label{sec:nbody}
In the \textsc{Fargo} and \textsc{Fargo3D} codes, the indirect term is
calculated at the beginning of the time step
and then added to the potential, which is then derived numerically to calculate the
acceleration on the gas. Then the source terms due to pressure, viscosity, and
artificial viscosity, and finally, the transport step are calculated and applied.

In this section, we construct a pure N-body test case
where only gravitational forces are considered, using the \textsc{Rebound} code
\citep{rein2012rebound}. This is a model of the full action to study the effects
of different protocols for including the indirect term.

We initialized a binary star system in \textsc{Rebound}
and shifted it into the frame of the primary star. We then updated the 
velocities of the N-body in time with the indirect term using a forward
Euler step in time before applying the standard \textsc{Rebound} integration step.
We used the \textit{IAS15} integrator \citep{rein2015ias15},
which is accurate to machine precision. We therefore assume that any
motion of the central object is due to inaccuracies in the indirect term protocols.
This allows us to measure their quality by how little the central object moves.

For a binary star system with masses $m_1,\,m_2$ and positions $\bf{x}_1,\,\bf{x}_2$,
the indirect term is simply given by the negative of the acceleration of the
primary by the secondary:
\begin{equation}
	\label{eq:euler}
	a_\mathrm{Ind,1} = m_2\frac{\bf{x}_1 - \bf{x}_2}{|\bf{x}_1 - \bf{x}_2|^3}.
\end{equation}
This formulation is identical to the implementation in the \textsc{Fargo} code.
We call it the Euler protocol because the velocity update is similar to an
Euler integrator in the sense that it updates the velocity with a single step forward
in time using the positions at the beginning of a time step.

As an alternative, we propose to use the acceleration experienced by the primary during its entire
update step to compute the indirect term. In our case, we extract the acceleration
on the central object from the \textsc{Rebound} code by measuring the change
in velocity during the entire integration step:
\begin{equation}
	\label{eq:rebound}
	a_\mathrm{Ind,2} = - \frac{{\bf v}_{1}(t + \Delta t) - {\bf v}_{1}(t)}{\Delta t}
\end{equation}
where $\Delta t$ is the size of the time step and ${\bf v}_{1}(t)$ is the velocity
of the central object at the time $t$.

To obtain the indirect term at the beginning of the time step, we create a 
copy of the N-body system and then integrate the copy with all the external forces acting on it
to compute the accelerations. Assuming that the velocity of the central object
at the start of the time step is zero, which is satisfied in a frame centered on it,
\eqref{eq:rebound} simply represents a velocity shift to the frame of the central object.
For this reason, we call this method the shift protocol for the indirect term.
We also computed the acceleration using a fifth-order Runge-Kutta scheme and found
identical results up to the accuracy of the integration scheme.

For our test, we used a circular binary system with a mass ratio of $m_2 / m_1 = 0.1$
and chose the units so that the total mass,
binary separation, and gravitational constant are all equal to one. With this setup,
one binary period is $P_\mathrm{bin} = 2 \pi$, which is divided into 250 steps
$\Delta t = 2 \pi / 250$ for integration. The primary star was initially placed in the
coordinate center.

\begin{figure}
	\centering
	\includegraphics[width=0.5\textwidth]{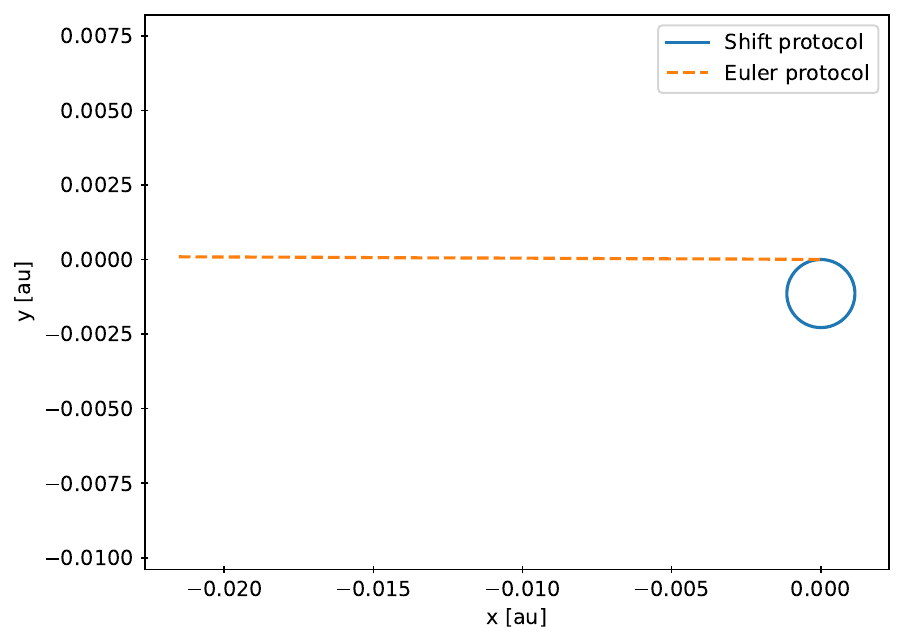}
	\caption{\label{fig:primary} Time evolution over 3 binary periods of the primary
  under the influence of the indirect terms protocols of~\eqref{eq:euler} and~\eqref{eq:rebound}.}
\end{figure}

The time evolution of the primary star in simulations using the Euler and shift
protocols for considering the indirect term are shown in \figref{fig:primary}.
Neither implementation keeps the primary fixed at ${\bf}x = 0$. For the shift
protocol, the velocity is zero at the end of each time step and the primary
drifts on a circle. For the Euler protocol, the primary reaches
zero velocity only after each full binary orbit and moves away from the origin.
By design, the shift protocol keeps the primary at zero velocity.
Any motion is only due to drift during the integration step as the primary
is accelerated from $-\Delta t a_\mathrm{Ind,2}$ to $0$.

For the next test, we shifted the binary system into the primary frame at the end of
each time step to fix the primary at the coordinate center.
This is intended to simulate the implementation of the \textsc{Fargo}
codes, in which the central object is implicit, i.e. it is not itself integrated and
the other N-body objects feel the indirect term at each substep of a fifth-order scheme,
so that they are held in the frame of the central object with high accuracy.

We then added massless test particles that were integrated alongside the binary. Their velocities
were updated with the indirect term at the beginning of the time step by adding $\vec{a}_\text{Ind} \,\Delta t$,
but were not shifted into the primary frame at the end of each time step.
They represent gas cells on a fixed grid and therefore cannot be
shifted to the primary frame.
We also added a test particle that is shifted to the primary frame along with the binary
after each time step as a reference.

We modeled the time step criteria of a hydrocode by scaling the time step by the inverse velocity
of the innermost particle, normalized by the 
Kepler velocity for a circular orbit at the same semi-major axis:

\begin{equation}
	\label{eq:cfl}
	\Delta t = \frac{2 \pi}{250} \cdot \frac{v_\mathrm{k}}{v_\mathrm{test,\, 1}}\;,
\end{equation}
which mimics the CFL criteria of the \textsc{Fargo} code.

\begin{figure}
	\centering
	\includegraphics[width=0.5\textwidth]{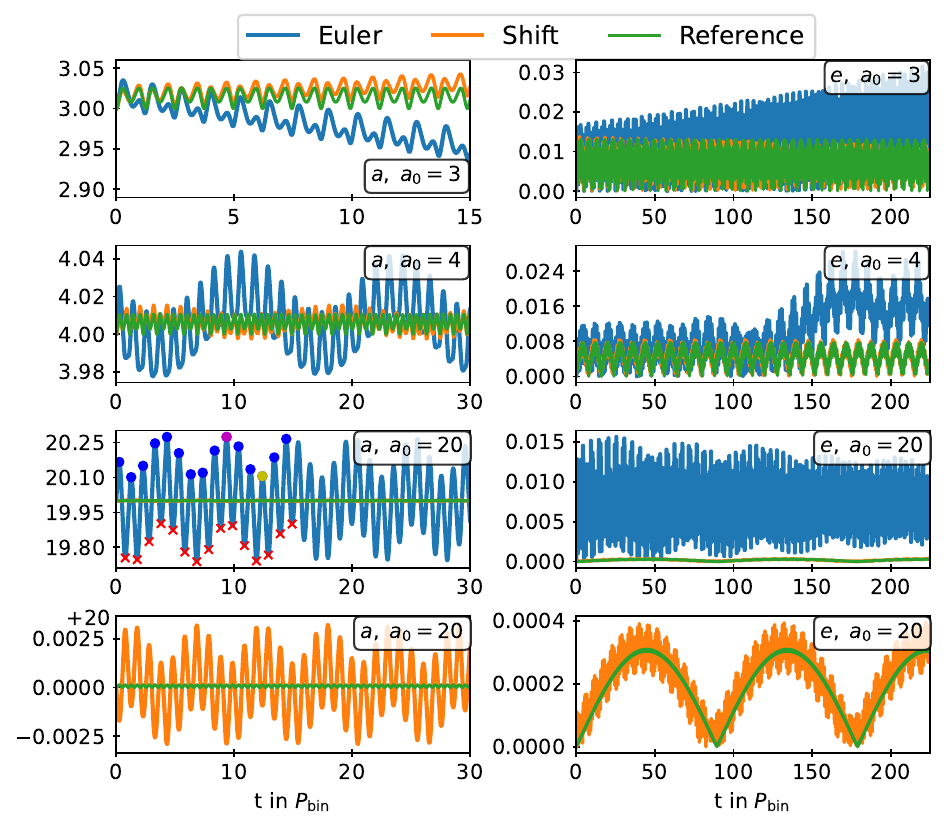}
	\caption{\label{fig:test_particles} Time evolution of the semi-major axis and eccentricity
  of two test particles under the influence of the indirect term in~\eqref{eq:euler}
  and~\eqref{eq:rebound} respectively and an exact reference case without indirect term.}
\end{figure}

The resulting evolution of the semi-major axis
and eccentricity for test particles starting at $3, 4, 20\,a_\mathrm{bin}$ are shown in
\figref{fig:test_particles}.
The first three rows correspond to a different initial semi-major axis,
$a_0$, and the bottom row shows a zoom in for the case of $a_0=20\,a_\mathrm{bin}$.
The left and right columns of the panels show the semi-major axis, $a$,
of the test particles and their eccentricity, $e$, as a function of time,
respectively. Note that the two columns show the quantities over
different time frames.

In total, we identified four different oscillations in the particle orbits.
The first has a period equal to the synodic period of the binary.
They can be seen in the semi-major axis evolution of the innermost
particle starting at $3~a_\mathrm{bin}$ in the upper left panel
in \figref{fig:test_particles}.
They are characterized by their double peaks and valleys during
a binary period as the particles move through the potential of the binary.
This oscillation is always present for the reference 
particle and is visible in the semi-major axis and eccentricity.
In the simulations with an indirect term, this oscillation 
is dominated by other oscillations and is no longer visible at larger radii.

The second oscillation has a frequency of the orbital period of the particles,
this oscillation occurs only in the eccentricity of the particles and 
it is clearly visible in the lower right panel. For reference,
the particles at $20\,a_\mathrm{bin}$ have a period of
$\approx 90\,P_\mathrm{bin}$.

In addition to these two physical oscillations, we find two purely numerical
oscillations for the particles with an indirect term,
which do not occur for the reference particle.
The first numerical oscillation also has a period equal to the synodic 
period of the binary, but produces only a single peak and valley 
instead of the two for the oscillation due to the binary potential.
We associate this oscillation with the residual motion due to the indirect
term protocols, similar to those shown in~\figref{fig:primary} for the central object.
Examples of this oscillation can be seen in the semi-major axis evolution
of the particles starting at $20\,a_\mathrm{bin}$ (third and fourth panels 
on the left in \figref{fig:test_particles}). They also appear in the eccentricity
of the particles, but cannot be seen individually because the eccentricity 
plots cover a larger time frame.

Finally, in simulations with indirect terms, the outer particles also oscillate
due to the feedback from the time-stepping criteria
which is calculated from the velocity of the innermost particle, see~\eqref{eq:cfl}. 
The frequency of this oscillation is the beat frequency between the 
orbital frequency of the particle itself and the innermost particle.
This CFL oscillation causes the height of the peaks and valleys to oscillate
during the synodic period, as seen in the semi-major axis and eccentricity
for particles starting at $4\,a_\mathrm{bin}$ and $20\,a_\mathrm{bin}$.
%
%
%

In addition to the oscillations, we observe other differences between the 
reference particles and the particles subject to an indirect term.
For the Euler protocol, the innermost particle (top row
of \figref{fig:test_particles}) drifts inward, and its eccentricity increases.
For the shift protocol, the particle drifts outward at one-third the drift of the
of the Euler protocol particle, and its eccentricity does not increase.

The sign of the drift direction and the sign
of the deviations from the reference particle in general, are reversed by
applying the indirect term after the integration step instead of before.
The drift of the innermost particles is caused by the variable
time step and does not occur with a constant time step.

The second row shows the time evolution of the second particle,
which started at a distance of $4\,a_\mathrm{bin}$.
The particles subject to the indirect terms are no longer drifting.
At $t = 150\,P_\mathrm{bin}$ the time step variations induce an eccentricity growth
in the simulation using the Euler protocol. We ran an additional simulation
where the time step from the Euler protocol simulation was
copied and used for the outer particle with the
shift-based protocol starting at $4~a_\mathrm{bin}$ (not shown),
which also caused it to increase its eccentricity, but at half the rate
of the case where both particles used the Euler protocol. 

The third row shows the particle at a distance of $20~\mathrm{au}$,
and the fourth row shows the same data again, but without the particle affected
by the Euler protocol. For the Euler protocol,
the eccentricity is higher by a factor of 35 and oscillates at twice the rate
of its binary period. The eccentricity of the shift protocol particle 
has the oscillation with its binary period 
as the dominant oscillation and still resembles the time evolution
of the reference particle.

We repeated the test and measured the amplitudes of the oscillations
in the semi-major axis of the outer test
particle for different binary mass ratios, time step sizes, and
initial semi-major axes.

An example of how we measured the oscillation amplitudes is shown in 
the third panel on the left in \figref{fig:test_particles}. We measured the amplitude
of the combined physical and shift protocol oscillations during
the synodic period, $A_\mathrm{syn}$,
as the differences between the maxima (blue dots) and the subsequent minima
(red crosses) of the semi-major axis.

\begin{figure}
	\centering
	\includegraphics[width=0.5\textwidth]{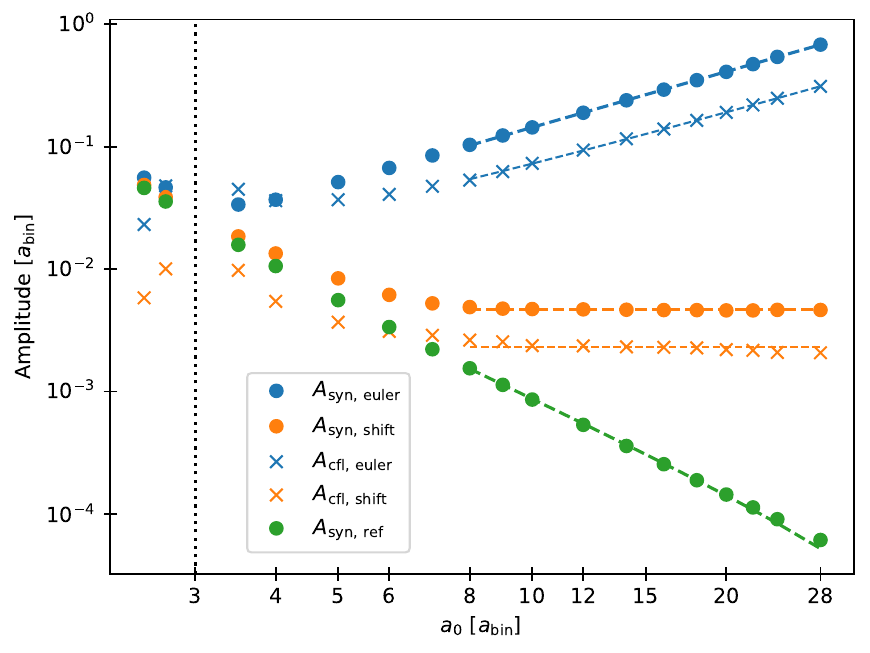}
	\caption{\label{fig:r_amplitudes} Amplitudes of the semi-major axis oscillations
	as a function of the initial semi-major axis. The dotted line at $3~a_\mathrm{bin}$ indicates
	the initial position of the 
	particle that is used for the time step criteria. The dashed lines represent the best least squares
	fit with a power law model.}
\end{figure}

We define the amplitude of the oscillations due to the CFL criteria,
$A_\mathrm{CFL}$, as the difference between the highest and lowest
maximum of the semi-major axis, as indicated by the purple and yellow dots.
The amplitudes were always measured during the first 30 binary
orbits, so that the particles were still close to their initial positions.
We ignored the numerical drift and used the initial semi-major axis of the particle
for the evaluation.
The initial semi-major axis of the particle
used for the time step criteria was kept
at $3\,a_\mathrm{bin}$, while the other particle was varied from $2.5$ to $28\,a_\mathrm{bin}$.
The resulting amplitudes for the different indirect term protocols as a function of
the initial semi-major axis are given in \figref{fig:r_amplitudes}. 

We evaluated the trends of the amplitudes by fitting a power law model to
the data of the form:
\begin{align}
	A = A_0 + c_1 a_0^{n_1} \cdot c_2 q^{n_2} \cdot c_3 \Delta t^{n_3}\, ,
\end{align}
where $a_0$ is the initial semi-major axis of the particle,
$q = m_2 / m_1$ is the mass ratio of the binary, and $\Delta t$ is the time step size.
The power law indices $n_i$ are fixed to the values in \tabref{tab:errors}, while the 
constant $A_0$ and the coefficients $c_i$ are determined by a least squares fit.
For all fitted parameters, we find relative $2\sigma$ errors of less than $11\%$, 
which we interpret as confirmation of the power law indices given in \tabref{tab:errors}.
The fits for the radial dependencies are plotted in \figref{fig:r_amplitudes} as
dashed lines. The other power law dependencies for time step and binary 
mass ratio were determined in the same way.
\newcommand\T{\rule{0pt}{2.6ex}} 
\newcommand\B{\rule[-1.4ex]{0pt}{0pt}} 
\begin{table}
\centering
\caption{\label{tab:errors} Scaling of the oscillation amplitudes in semi-major axis
found in \figref{fig:r_amplitudes}
 for the indirect term protocols for particles beyond $8~a_\mathrm{bin}$.}
 \renewcommand{\arraystretch}{1.2}
\begin{tabular}{c c c c}
	\hline
	\hline
	\rule{0pt}{2.5ex} Protocol & $\frac{\mathrm{d\;log}\, A_\mathrm{syn}}{\mathrm{d\; log}\, a_0}$ & $\frac{\mathrm{d\;log}\, A_\mathrm{syn}}{\mathrm{d\; log}\, \Delta t}$ & $\frac{\mathrm{d\;log}\, A_\mathrm{syn}}{\mathrm{d\; log}\, q}$\T\B \\
	\hline
	Reference & -2.5 & 0 & 1\\
	Shift & 0 & 1 & 1\\
	Euler & 1.5 & 1 & 1\\
	\rule{0pt}{4ex}
	\rule{0pt}{2.5ex} & $\frac{\mathrm{d\;log}\, A_\mathrm{cfl}}{\mathrm{d\; log}\, a_0}$ & $\frac{\mathrm{d\;log}\, A_\mathrm{cfl}}{\mathrm{d\; log}\, \Delta t}$ & $\frac{\mathrm{d\;log}\, A_\mathrm{cfl}}{\mathrm{d\; log}\, q}$\T\B \\
	\hline
	Reference & - & - & -\\
	Shift & 0 & 1 & 2\\
	Euler & 1.5 & 1 & 2\\
	\hline
\end{tabular}
\end{table}

For both indirect term protocols, we find oscillation amplitudes that scale
linearly with the time step size, as expected from a first-order method.
The amplitudes of the synodic oscillations depend linearly on the mass ratio,
identical to the reference model, since they are physical and caused by the perturbations of the companion star,
which scale linearly with its mass.
The amplitudes of the oscillations due to the CFL criteria scale with the square 
of the companion mass, because they depend on the strength of the perturbation 
as well as on the velocity variations of the innermost particle according to~\eqref{eq:cfl}.

From the reference particles we infer that the amplitude of the synodic
oscillations due to the variations in the binary potential scale with the initial semi-major axis as $a_0^{-2.5}$.
For the shift protocol, the amplitude of the synodic oscillations
reaches a constant floor value at distances above $8~a_\mathrm{bin}$.
We interpret this as the oscillations being dominated by the potential 
close to the binary. This causes the 
particle using the shift protocol to behave similarly to the reference particle
at distances below $8~a_\mathrm{bin}$ while beyond $8~a_\mathrm{bin}$, the numerical
oscillations dominate and their behavior diverges. 
For the Euler protocol, the numerical oscillation amplitudes increase with $a_0^{1.5}$ and are already 
significantly larger than the reference amplitude at an initial
separation of $4~a_\mathrm{bin}$. 
%
%

For both the synodic period oscillation and the time step criteria oscillation,
the magnitude of the errors of the Euler protocol at large radii is a
factor of $(a_0 / a_\mathrm{bin})^{1.5}$ larger than the errors of the shift protocol.
Since $a_0 > a_\mathrm{bin}$ for circumbinary particles, 
the errors of the Euler protocol will always be larger than the
errors of the shift protocol in this case.

We repeated similar tests for particles on an s-type orbit,
that is, orbiting the central object with the companion as an outside perturber.
For this configuration, the indirect term will always be weaker than
the direct gravitational forces.
We used the same binary setup as before and put the inner particle used for
the time step criteria at $0.3~a_\mathrm{bin}$ and the outer particle at
$0.5~a_\mathrm{bin}$. 
We reduced the time step so that the orbit of the inner particle was
resolved with 250 steps. For this setup, both indirect term protocols
induced an artificial inward drift that scales linearly with the time step size,
with the shift protocol leading to a 20\% faster drift than the Euler protocol.
The deviations of the oscillation amplitudes from the reference particle 
are on the order of $10^{-3}$ and can be positive or negative depending
on the time step size for both protocols. Thus, we do not find 
a clear trend of increasing inaccuracies as we did for the p-type orbits before.

We conclude that single-step-forward-in-time methods for the indirect
term introduce inaccuracies that scale with the time step size
and the strength of the indirect term. When the time step 
is determined by the velocity of a particle, analogous to a cell in a hydrodynamic simulation,
it creates a feedback effect that causes the particle itself to migrate and 
induce oscillations in the other particles.
We have introduced a new method for calculating the indirect term that is 
more accurate and stable for p-type orbits (orbits around both binary stars) than the
standard indirect term protocol typically used in hydrocodes.
\section{Artificial viscosity}
\label{sec:art_vis}
The \textsc{Fargo} codes use a second-order upwind scheme for 
the hydrodynamic advection, which cannot handle discontinuities
and requires artificial  viscosity to spread shocks over multiple
grid cells. 
By default, \textsc{Fargo} codes use the artificial viscosity developed by
\citet{neumann1950shocks} (hereafter \citetalias{neumann1950shocks}), which acts as
a bulk viscosity and enters the equation of motion as an anisotropic artificial
pressure to counteract compression.
The implementation of the artificial viscosity is described in \citet[Section 4.3]{stone1992zeus2d}
(hereafter \citetalias{stone1992zeus2d}), who define the artificial viscosity
in one dimension and apply it to each dimension independently:
\begin{equation}
	\label{eq:art_SN}
	Q_i =
	\begin{cases}
		c^2 \Delta x_i^2 \Sigma \left(\frac{\partial v_i}{\partial x_i}\right)^2 & \text{if\;} \frac{\partial v_i}{\partial x_i} < 0 \\
		0                                      & \text{otherwise.}            \\
	\end{cases}
\end{equation}
where $c$ is the artificial viscosity constant that measures the number of cells
over which the shock spreads, for which values around 2 are typically recommended, and $\Delta x_i$
is the cell size along direction $i$.
Note the misprint in Eq.~(33) \& (34) by \citetalias{stone1992zeus2d}, where they forgot to square
the artificial viscosity constant $c$.

\citet{tscharnuter1979artificial_viscosity} (hereafter \citetalias{tscharnuter1979artificial_viscosity})
raised concerns about the formulation of
artificial viscosity by \citetalias{neumann1950shocks} being used on curve-linear
coordinate systems, since it is formally valid only for Cartesian coordinates.
They show that it can create artificial pressure in curve-linear coordinates
even when there are no shocks. A specific example is given 
in \citet[Section 6.1.4]{bodenheimer2006numerical} where they show that the
artificial  viscosity by \citetalias{neumann1950shocks} creates an
artificial pressure that accelerates the collapse of a free-falling
spherical gas cloud ($|v_r| \propto r$) even though the flow is smooth.

\citetalias{tscharnuter1979artificial_viscosity} proposed a
tensor artificial viscosity, analogous to the viscous stress tensor, which is independent
of the coordinate system and frame of reference. An implementation of this
artificial viscosity is also described in \citetalias[Appendix B]{stone1992zeus2d},
who added two additional constraints to the artificial
viscosity prescription by \citetalias{tscharnuter1979artificial_viscosity}:
the artificial viscosity constant must be the same in all directions (isotropy), and
the off-diagonal elements of the tensor must be zero to avoid artificial
angular momentum transport.
The resulting artificial viscosity pressure tensor is given by:
\begin{equation}
	\label{eq:art_TW}
	\mathbf{Q} =
	\begin{cases}
		c^2 \Delta x^2 \Sigma \nabla\mathbf{v}\left[\nabla\otimes\mathbf{v} - \frac{1}{3} \nabla\mathbf{v}\,\mathbf{I}\right] & \text{if\;} \nabla \mathbf{v} < 0 \,, \\
		0                                                                                                          & \text{otherwise,}                 \\
	\end{cases}
\end{equation}
where $c$ is again a dimensionless parameter near unity and $\Delta x$ is
the maximum cell extension in any direction.
%

We can analytically estimate the effect of the more naive approach given
in~\eqref{eq:art_SN} for a disk in a non-inertial frame.
We consider the case of a Keplerian gas disk around a circular two-body system in the center-of-mass frame,
where the primary moves at a velocity of $\mathbf{v}_0$.
If the radial velocity of the gas is neglected, the gas will have the Cartesian velocity
components:
\begin{equation}
v_x = -v_\mathrm{k} \sin(\phi)\\ \text{and} \\
v_y = v_\mathrm{k} \cos(\phi)\;,
\end{equation}
where $v_\mathrm{k}$ is the Keplerian velocity and $\phi$ is the angle in the center-of-mass frame.
We then shift the whole system to the
primary frame by subtracting its position and velocity from the gas and binary.
After shifting to the primary center, the gas has the following velocity components:
\begin{eqnarray}
v_r = v_\mathrm{k} \sin(\varphi - \phi) + v_{0} \sin(\varphi - \varphi_0)~,\\
v_\varphi = v_\mathrm{k} \cos(\varphi - \phi) + v_{0} \cos(\varphi - \varphi_0)~,
\end{eqnarray}
where $\varphi$ is the angle in the primary frame and $\varphi_0$ is the azimuthal angle of the companion in the primary frame.
Without loss of generality, we assume $\varphi_0 = 0$, meaning that the binary is located on the x-axis
and the secondary is moving in the positive y-direction in the primary frame.
In addition, we make the simplifying assumption that the length of the position
vector of the cell is much greater than the length of the shift vector from the center of mass to the primary,  
so that its angular position is the same in the center-of-mass frame and in the primary frame: $\phi \approx \varphi$.
Under these assumptions, the velocity of the gas has no radial dependence.
This causes the radial component of the artificial pressure in \eqref{eq:art_SN},
which depends only on the radial derivatives of the radial velocity, to always be zero.
Therefore, we do not consider the radial velocity any further.
The azimuthal velocity becomes:
\begin{equation}
\label{eq:v_phi}
v_\varphi = v_\mathrm{k} + v_{0} \cos(\varphi)~.
\end{equation}
Putting this velocity into~\eqref{eq:art_TW} results in zero artificial viscosity,
while putting it into~\eqref{eq:art_SN} results in an artificial viscosity of:
\begin{equation}
	\label{eq:artificial_pressure}
	Q_\varphi =
	\begin{cases}
		c^2 \Sigma \left[- \Delta \varphi \cdot v_0  \sin(\varphi) \right]^2 & \text{if\;} 0 < \varphi < \pi \\
		0                                      & \text{if\;} \pi < \varphi < 2\pi            \\
	\end{cases}\;,
\end{equation}
where $\Delta \varphi = 2\pi / N_\varphi$ is the angular cell size of the grid
and $N_\varphi$ is the number of azimuthal grid cells.
The acceleration due to the artificial viscosity is then:
\begin{eqnarray}
	\label{eq:artificial_force}
	\frac{\partial v_\varphi}{\partial t}
	= -\frac{1}{\Sigma} \frac{1}{r} \frac{\partial Q_\varphi}{\partial \varphi}
	= 
	\begin{cases}
		-\frac{1}{r} c^2 \Delta \varphi ^2 v_0^2 \sin(2\varphi) & \text{if\;} 0 < \varphi < \pi \\
		0                                      & \text{if\;} \pi < \varphi < 2\pi\;,
	\end{cases}
\end{eqnarray}
If we use units for which $\mathrm{G}M = 1$, and assume that the cell is far enough away
from the two-body object that
the distance to the primary at the center is approximately equal to the
distance to the companion,
we find the ratio of the artificial acceleration, $\dot{v}_\mathrm{\varphi,~art}$, to the magnitude of the direct
gravitational acceleration caused by the companion, $|\mathbf{a}_2|$, to be:
\begin{eqnarray}\label{eq:artificial_force_ratio}
	\frac{\dot{v}_\mathrm{\varphi,~art}}{|\mathbf{a}_2|} = c^2 \Delta \varphi^2 \frac{r}{a} \frac{q}{(1+q)} \sin(2\varphi)\;,
\end{eqnarray}
where $q$ is the mass ratio of the companion to the primary and $a$ is the semi-major axis.
For typical values used in planet-disk interaction simulations with
a Jupiter-mass planet, e.g.
$c = \sqrt{2}$, $q = 10^{-3}$, $r = 10\,a$, and $N_\varphi = 1024$,
we find a ratio of the artificial acceleration to the acceleration caused by the planet of
up to $7.5\cdot 10^{-7}$.
Assuming a locally isothermal equation of state and a scale height of $h=0.05$,
the maximum artificial pressure due to~\eqref{eq:artificial_pressure} 
is $3\cdot 10^{-7}$ times the pressure of the gas.

\begin{figure}
	\begin{center}
		\includegraphics[width=\linewidth]{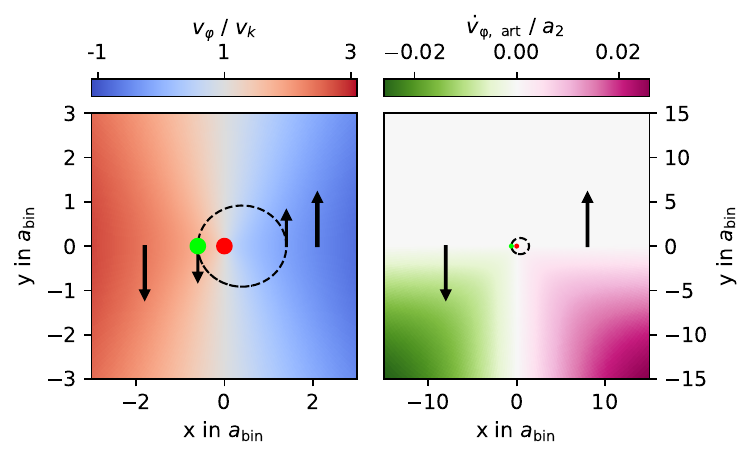}
		\caption{Snapshot of the azimuthal velocity (left) and artificial acceleration 
		due to~\eqref{eq:artificial_force_ratio} in the frame of the secondary (right).
		The red dot represents the secondary, the green dot represents the primary,
		and the black dashed line represents the orbit of the primary.
		The arrows indicate the direction of motion and are not to scale.
		The binary and grid parameters are the same as in \secref{sec:binary}.
		\label{fig:art_accel}}
	\end{center}
\end{figure}

For a circumbinary disk, this ratio becomes much larger and very relevant in the simulations.
The artificial accelerations due to~\eqref{eq:artificial_force_ratio} 
 are shown in \figref{fig:art_accel} for a binary system in the frame of the secondary.
The simulation parameters are the same as in \secref{sec:binary}, the binary has a mass ratio of $q = 2$
and an eccentricity of $e_\mathrm{bin} = 0.4$, the grid has $N_\varphi = 315$ azimuthal cells, 
and the artificial viscosity constant is $c = \sqrt{2}$. The binary is currently in the pericenter.

For the parameters of this model, our estimate for the artificial pressure
using~\eqref{eq:artificial_force_ratio} predicts a force during the binary 
periastron passage that is $0.6\%$ of the gravitational force of the primary
at $r=5\,a_\mathrm{bin}$, which is two-thirds of the pressure forces of the gas itself.

The left panel is zoomed in on the binary and depicts the azimuthal velocity of the gas
according to \eqref{eq:v_phi} normalized by the Keplerian velocity. It is shown to visualize 
the setup and its extreme case where the velocity of the gas is dominated by the velocity 
of the frame.
The right panel depcits the resulting acceleration field due to the SN artificial viscosity 
normalized by the gravitational acceleration caused by the primary (Eq.~\ref{eq:artificial_force_ratio}).
The right panel of \figref{fig:art_accel} covers the same area as \figref{fig:indirect_term}.

The depicted acceleration field rotates with the binary.
Looking at a fixed cell far from the binary 
(with a slower orbital frequency than that of the binary) as the acceleration field rotates
with the binary, the deceleration and subsequent acceleration result in an epicyclic motion 
with the same frequency as the orbital frequency of the binary.
In a non-isothermal simulation, the artificial viscosity in~\eqref{eq:artificial_pressure}
would be used to calculate the shock heating of the gas and the gas would be artificially heated.

If the binary is eccentric, then the accelerations during a binary orbit 
would not cancel each other out, instead the acceleration during the binary periastron passage
would be higher than during the apastron passage. This leads to a 
residual acceleration field with the same shape as the one presented in \figref{fig:art_accel}.
The deceleration and acceleration during an orbit is a similar pattern to a particle
on an eccentric orbit. In this case, the artificial acceleration could excite an
eccentricity with a preferred direction based on the location of the binary periastron.

A circumplanetary disk in the frame of the star will have similar problems.
The circular motion around the planet will appear as radial and azimuthal
velocity components oscillating around the planet in the frame of the central star.
The resulting radial and azimuthal velocity gradients 
are treated as shocks by the artificial viscosity
prescription in~\eqref{eq:art_SN}. In addition, 
the resolution in the vicinity of a planet on a logarithmic polar grid
is typically much lower than the global resolution.
We analyze the influence of the different choices of artificial viscosity on a circumplanetary disk further in Appendix~\ref{app:companion_disk}.
In our tests presented in \secref{sec:planet},
we find that the disks around the companion begin to behave differently
for the two artificial viscosity prescriptions at companion
masses of $3~M_\mathrm{Jup}$ ($q \geq 3\cdot 10^{-3}$).

\section{Circumbinary disk in non-inertial frame}
\label{sec:binary}
To highlight the importance of our changes to the indirect term and the use of
the artificial viscosity by \citetalias{tscharnuter1979artificial_viscosity}, we constructed
a test by simulating a circumbinary disk in the center-of-mass frame and in the center
of the secondary star. For all our hydrodynamical simulations, we use the \textsc{FargoCPT} 
code, which is freely available on GitHub\footnote[1]{\url{https://github.com/rometsch/fargocpt}}  and described in detail in 
\citet{rometsch2024fargocpt}.

The simulations are locally isothermal and we use the $\alpha$ viscosity prescription from
\citet{shakura1973alpha} ($\alpha=0.01$, $h=0.04\cdot (d/a_\mathrm{bin})^{0.3}$, $GM=1$).
The surface density is initialized as $\Sigma(d) = \Sigma_0 \cdot (d/a_\mathrm{bin})^{-1.1}$.
The gas velocities and surface densities are always initialized in the center-of-mass frame,
and $d$ is the distance to the center of mass, which is not equal to the $r$ coordinate of
the grid in the secondary frame.
The setup is intended to model a steady state accretion disk around a binary star system.
The artificial viscosity constant is set to $c = \sqrt{2}$.

The domain is logarithmically spaced from 1 to $60\,a_\mathrm{bin}$
in the radial direction and uniformly spaced from $0$ to $2\pi$ in the azimuthal direction
with a resolution of $\mathrm{N_r} \times \mathrm{N_\varphi} = 207 \times 315$,
or equally, 2 cells per scale height at $r = 1\,a_\mathrm{bin}$. The resolution 
is chosen low to make this a stress test for the numerical scheme.
The disk is damped to the initial conditions from 36 to $60\,a_\mathrm{bin}$
on a timescale of $10^{-3}$ Keplerian orbital periods at $60\,a_\mathrm{bin}$.
We applied strict outflow conditions at the inner boundary, and in case 
a star enters the simulation domain, we added a sink hole around each star,
which removes a fraction of the gas within its Roche lobe with a half-emptying time of $10^{-2}\,T_\mathrm{orb}$.
The removed mass is included in the accretion rate, but is not added to the mass of the stars.
We used a density floor of $10^{-7} \cdot \Sigma_0$ for numerical stability, meaning that if 
the density of a cell falls below this value, it will be set to this value.
The gravity of the gas is ignored. The mass of the secondary is $m_\mathrm{Sec} = 0.5$
and $m_\mathrm{Prim} = 1$ for the primary star. The binary eccentricity is
$e_\mathrm{bin} = 0.4$, which causes the primary to enter the simulation domain 
during its orbit when the simulation is centered on the secondary.
To prevent this from causing numerical problems, we smooth the gravitational potential of the 
primary within a radius of $0.5\,R_\mathrm{Hill}$ around it using the
third-order polynomial prescription by \citet[][Eq.~4]{klahr2006}.
To make the time step size comparable between the simulations, we use a Courant number of $0.4$
in simulations using the artificial viscosity by \citet{stone1992zeus2d} and a 
Courant number of $0.5$ in simulations using the artificial viscosity by \citet{tscharnuter1979artificial_viscosity}.

We ran the setup with different combinations of coordinate centers
(default is centered on the secondary
and 'CMS' indicates the simulation is in the center-of-mass system)
and artificial viscosity ('TW' is the artificial
viscosity by \citetalias{tscharnuter1979artificial_viscosity}  and 'SN' is the artificial
viscosity by \citetalias{neumann1950shocks}, implemented as described in \citetalias{stone1992zeus2d}).
The 'CMS' simulations are a reference case for the indirect term
experiments, since the indirect term vanishes in the center-of-mass system.
\begin{figure}
	\begin{center}
		\includegraphics[width=\linewidth]{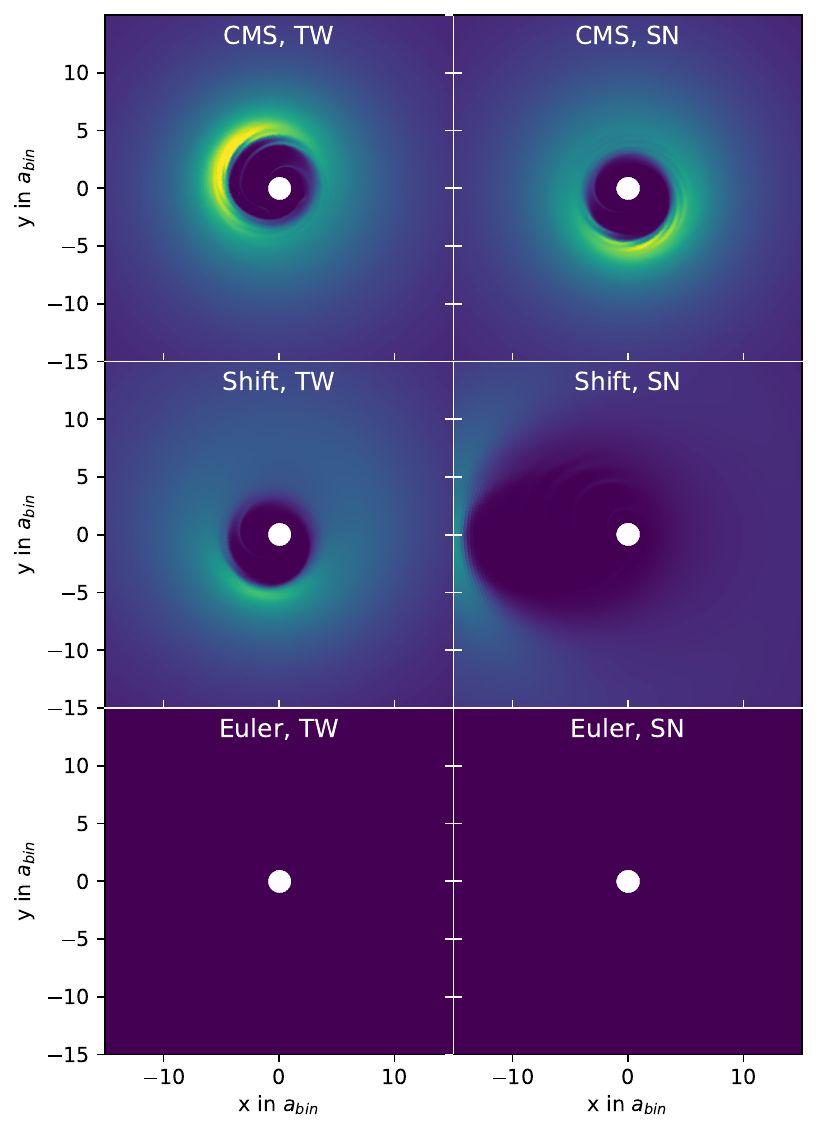}
		\caption{Snapshot of the surface density after $5000$ binary orbits of a massless disk
		around a binary system with $q_\mathrm{bin} = 0.5$ and $e_\mathrm{bin} = 0.4$.
		The plots show the results for different combinations of coordinate centers,
		artificial viscosity, and indirect term protocol.
		All plots use the same linear color scale.
		\label{fig:indirect_term}}
	\end{center}
\end{figure}

Snapshots of the surface densities after $5000~T_\mathrm{bin}$ are shown in
Fig.\,\ref{fig:indirect_term}.
The top row shows the simulation in the center-of-mass frame,
where there are no indirect forces.
For both artificial viscosities, the disks become eccentric and precess
in the prograde direction, which is the expected behavior for a
circumbinary disk \citep{duffel2024}.
By evaluating the disk quantities averaged between $t = 5000-8000~T_\mathrm{bin}$, we find 
that compared to TW, the disk simulated with the SN artificial viscosity 
has a $13~\%$ faster precession rate, $9~\%$ higher eccentricity,
$2~\%$ larger gap, and $12~\%$ lower peak density at 
longitude at apoastron. While these differences are significant, 
we expect them to become negligible when using an appropriate grid resolution
(higher than $8$ cells per scale height). 

In the second row, the grid of the hydrodynamics simulation is centered on the 
secondary star of the binary and the shift-based indirect term protocol is used.
Here, the choice of the artificial viscosity is more significant.
For the TW artificial viscosity, the inner gap becomes eccentric and precesses, 
resembling the CMS simulations.
However, the precession time is roughly a factor of 4 longer than in the CMS simulations, 
the densities in the ring outside the gap are lower and the spiral patterns inside the density 
ring do not form.

For 'SN' artificial viscosity, the simulation evolves like the others at the start,
but instead of precessing, the longitude of apoastron (eccentric bulge) of the gap 
remains at $\varphi = \pi / 2$ (the negative y-axis). At about $T=1500~T_\mathrm{bin}$,
the eccentricity of the gap starts to grow, while the longitude of apoastron
moves towards $\varphi = \pi$ (negative x-axis), which is also the position of the periastron
of the binary, and stays there.
The indirect term is the strongest towards the pericenter of the binary ($\varphi_0 = \pi$).
If we substitute $\varphi \rightarrow \varphi - \varphi_0$ in~\eqref{eq:artificial_force},
we find that the gas is decelerated from $\varphi = \pi$ to $\varphi = 3/2 \pi$ and accelerated
from $\varphi = 3/2 \pi$ to $\varphi = 2\pi$. The gas thus has the maximum velocity due to 
the artificial acceleration at the x-axis ($\varphi = 0$), compare \figref{fig:art_accel}. 
This is in consistent with the preferred position of the longitude of pericenter of the eccentric disk
that we find in the simulation.
%
At the end of the simulation at $8000\,T_\mathrm{bin}$, the gap edge extented 
to $13\,a_\mathrm{bin}$ compared to the $4.5\,a_\mathrm{bin}$ for the TW artificial viscosity
and also both the CMS simulations.

The eccentricity of the disk increases and decreases depending on 
the phase of the disk due to interaction with the binary potential.
Therefore, the eccentricity instability is caused by the SN artificial
viscosity preventing the disk from precessing and holding it
in a position where the eccentricity is increasing.
If the artificial forces are too weak to stop the precession, 
the artificial accelerations in \figref{fig:art_accel} cancel out over
a precession period, and the artificial viscosities can converge
at increasing resolutions.

In simulations using the Euler protocol, the disk completely dispersed within
the first 200 binary orbits.

We repeated the test for different resolutions.
When the resolution was reduced to $N_r \times N_\varphi = 104 \times 156$,
the shift-based, TW artificial viscosity simulation also became unstable.
When the resolution was doubled to $N_r \times N_\varphi = 411 \times 628$,
the simulation using the shift-based indirect term with the SN artificial 
remained stable for the entire simulation time of $7000~T_\mathrm{bin}$.
In both cases, the center-of-mass frame and the secondary frame, 
the SN artificial viscosity produced different precession rates 
and mass accretion rates through the inner boundary compared to the 
equivalent simulation using the TW artificial viscosity.

\begin{figure}
	\begin{center}
		\includegraphics[width=\linewidth]{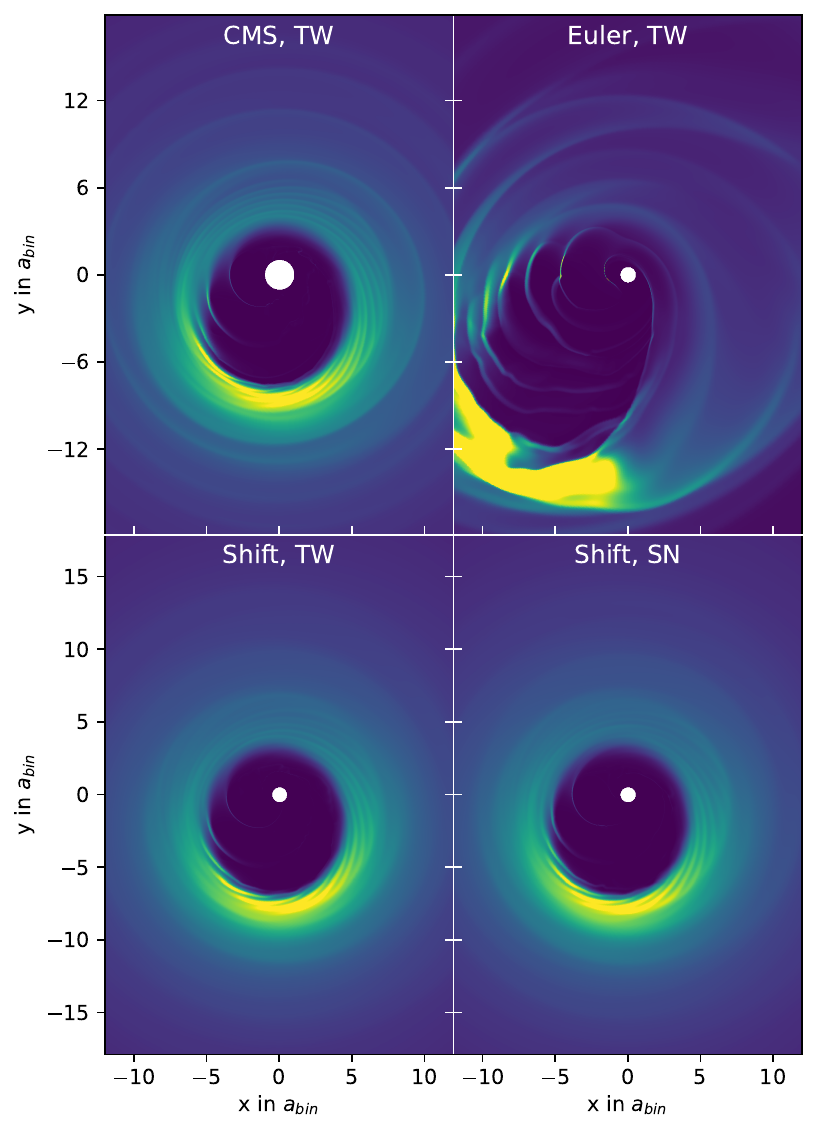}
		\caption{Snapshot of the surface densities of the high resolution
		simulations with a binary eccentricity of $0.4$.
		The simulation using the Euler indirect is plotted at a simulation time
		of $t=800~T_\mathrm{bin}$, while the other simulations are plotted at 
		$t=4000~T_\mathrm{bin}$. \label{fig:indirect_term_highres}
		All plots use the same linear color scale.
		}
	\end{center}
\end{figure}

The surface densities of our high resolution (12 cells per scale height) runs
are shown in \figref{fig:indirect_term_highres}.
An inner domain radius of $1~a_\mathrm{bin}$ is barely sufficient 
for a simulation in the center-of-mass frame \citep{thun2017numerical},
and in a star-centered simulation the inner boundary will be even closer to the
circumbinary disk. Therefore, we have reduced the inner domain radius to $0.5~a_\mathrm{bin}$
for the simulations that are centered on the secondary, leading to a grid resolution of $1439\times1885$.

Even at this resolution, the disk becomes unstable
right at the beginning of the simulation if the Euler protocol is used,
as shown in the top right plot in \figref{fig:indirect_term_highres}.
The highly eccentric disk does not precess and remains
in the depicted orientation. In a previous run with a larger inner domain radius,
the disk aligned with the positive y-axis, so we do not find a 
preferred orientation for this instability.
Note that the instability here
is caused by the indirect term protocol, as opposed to 
the instability caused by the artificial viscosity we found at lower resolutions,
which leads to an alignment with the negative x-axis.

\begin{figure}
	\begin{center}
		\includegraphics[width=\linewidth]{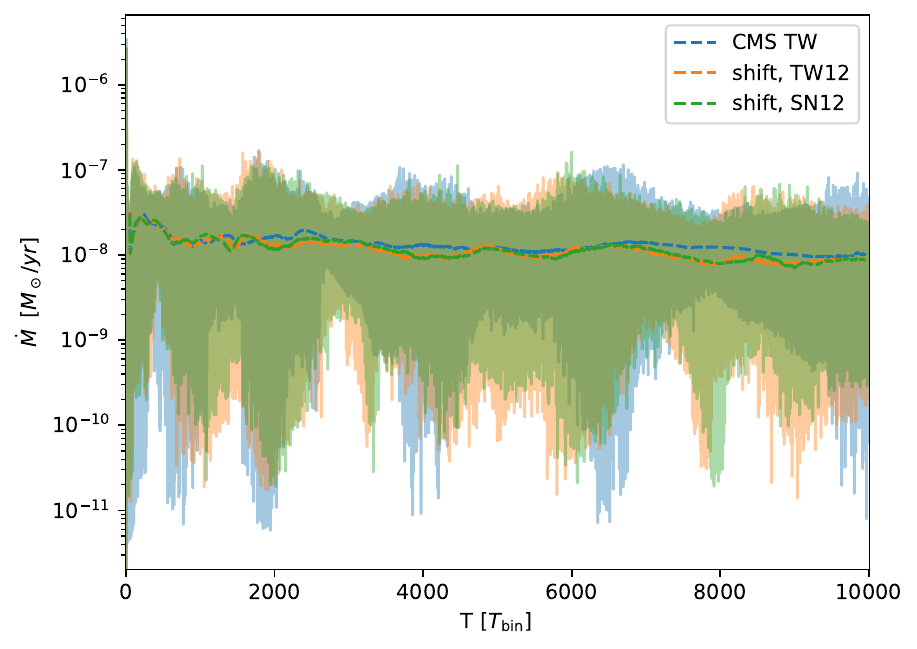}
		\caption{Mass accretion rate of the circumbinary disk through the inner boundary over time.
		The dashed lines show the moving average over 100 binary orbits.
		\label{fig:mdot_highres}}
	\end{center}
\end{figure}

With the shift protocol, the disks in the secondary frame 
produce the same gap profile as the CMS reference case 
and also precess.
The similarity of the dynamic of the disks can also be seen by the 
similarity of the mass accretion rates shown in \figref{fig:mdot_highres}.
At the start of the simulations, the mass accretion rates are nearly identical 
and then slowly drift apart due to differences in indirect term and 
artificial viscosity.
Compared to the CMS simulation, the SN artificial viscosity underestimates
the mass accretion rate by $11.4\%$ and the TW artificial viscosity by $12.5\%$.
The lower mass accretion rate can be partly attributed to the smaller inner
domain radius, but also to different reference frame.
We have also conducted simulations with even smaller inner radii, which are not shown here,
where the inner radius was contained within the Roche lobe of the 
stars for both setups, and still found lower mass accretion rates in the star-centered simulation.

From \eqref{eq:artificial_force_ratio} we estimate that the 
artificial pressure forces caused by the SN artificial viscosity 
are $2\%$ of the pressure forces of the gas, which is 
consistent with the small deviations we observe in \figref{fig:mdot_highres}.
The accuracy of the indirect term could be further improved
by using smaller time steps, which 
automatically occurs when the radius of the inner domain is reduced 
to include the circumstellar disks.


\section{Protoplanetary disk with hot jupiter}
\label{sec:planet}
In this section, we test whether our proposed changes are relevant to more common scenarios
of planet-disk interactions. We use a setup of a central star of mass $m_\mathrm{star} = 1 M_\odot$
and a companion object with masses of $m = [1, 3, 10, 50] M_\mathrm{Jup}$
(corresponding to mass ratios of $q\approx 0.1\%, 0.3\%, 1\%, 5\%$) on a circular orbit ($e=0$) with an initial semi-major axis of $1\,\mathrm{au}$.
At the beginning of the simulation, the companion is
on a fixed orbit and is treated as massless when interacting with the disk. 
The mass of the compaion felt by the gas is then ramped up over the first 50 orbital periods according to
\begin{equation}
	\label{eq:mass_growth}
	m(t) = \begin{cases}
	m \cdot \left[1.0 - \cos^2(t \cdot 2\pi / 50 T_\mathrm{orb})\right], \;\;\; t < 50 T_\mathrm{orb}
	\\
	m, \;\;\;\;\;\;\;\;\;\;\;\;\;\;\;\;\;\;\;\;\;\;\;\;\;\;\;\;\;\;\;\;\;\;\;\;\;\;\;\;\;\;\;\; t \geq 50 T_\mathrm{orb}.
	\end{cases}
\end{equation}
Despite the mass ramping, the gas velocities were always initialized in the center-of-mass frame 
of the N-body system with their fully ramped-up masses.

The gravitational potential of the N-body system at the position of a cell is computed as a 
Plummer potential with a smoothing length of $\epsilon = 0.6 H$ \citep{mueller2012thin},
where $H = h\cdot r$ is the gas scale height. The potential reads:
\begin{equation}
	\label{eq:potential}
	\Phi_i  = -G\sum_{k}^{N_\mathrm{nb}} \frac{M_k}{s_{ik}}\;,
\end{equation}
where $G$ is the gravitational constant, $s_{ik} = \sqrt{d_{ik}^2 + \epsilon^2}$,
is the smoothed distance between the gas cell $i$ and the N-body object
$k$ with mass $M_k$.

Similarly, the force exerted by the gas on the N-body objects is calculated as
\begin{equation}
	\label{eq:force}
	\mathbf{a}_k  = -G \sum_{i}^{N_\mathrm{cell}} f_\mathrm{sm}(s_{ik}) \frac{m_i}{s_{ik}^3} \mathbf{d}_{ik}\;,
\end{equation}
where $m_i$ is the mass of the gas cell and $f_\mathrm{sm}$ is the smoothing
function by \citet{crida2009smooth}:
\begin{equation}
	\label{eq:smooth}
	f_\mathrm{sm}(s)  = \left[\exp\left(-10\left(\frac{s}{0.8\,r_H}-1 \right) \right) +1 \right]^{-1}\;,
\end{equation}
where $r_H$ is the Hill radius according to \citet{eggleton1983roche_radius}. The smoothing is applied only
to the companion object, not to the star. It acts as a filter 
to remove the effects of a disk around the companion, since the disk is poorly resolved
and not realistically modeled in our setup.

The disk is locally isothermal ($\alpha=10^{-3}$, $h=0.05$).
The simulation domain ranges from $0.25\,\mathrm{au}$ to $25\,\mathrm{au}$ 
and the cells are spaced logarithmically in the radial direction and 
uniformly in the azimuthal direction
with a resolution of $\mathrm{N_r} \times \mathrm{N_\varphi} = 739 \times 1005$ which 
corresponds to 8 cells per scale height.
The surface density is initialized as
\begin{multline}
	\Sigma(d) = 200\,\mathrm{g\,cm^{-2}} \cdot
	(d/\mathrm{1 au})^{-0.5} \cdot \frac{1}{1 + \exp[(d-12\,\mathrm{au})/0.25\,\mathrm{au}]}~,
\end{multline}
where $d$ is the distance to the center-of-mass of the N-body system.
We applied strict outflow boundary conditions
at the inner and outer boundaries. Since the location of the boundaries depends on the frame of reference,
we added an exponential cutoff at $12\,\mathrm{au}$ to prevent the disk 
from interacting with the outer boundary instead of applying damping conditions in the outer regions.

We measured the accretion on the companion using the accretion model of
\citet{kley1999accretion}, that is, we removed a fraction of the gas from 
the Hill sphere of the companion at each time step with a half-emptying time of $1000\,T_\mathrm{orb}$.
The mass of the gas and its momentum were added to the companion.

The first 200 orbital periods are used for initialization, during which time
the disk exerts no force on the star or companion, and gas accretion
by the companion is not active. Then disk feedback and companion accretion
are activated, and the simulations run for another 1800 orbital periods.
The companion will then migrate under the gravitational influence of the disk
and due to momentum accretion. The gravitational force of the gas on the star
is added to the indirect term of the simulation.
The N-body system is always shifted to the frame of the star at the end of each time step.

\begin{figure}
	\begin{center}
		\includegraphics[width=\linewidth]{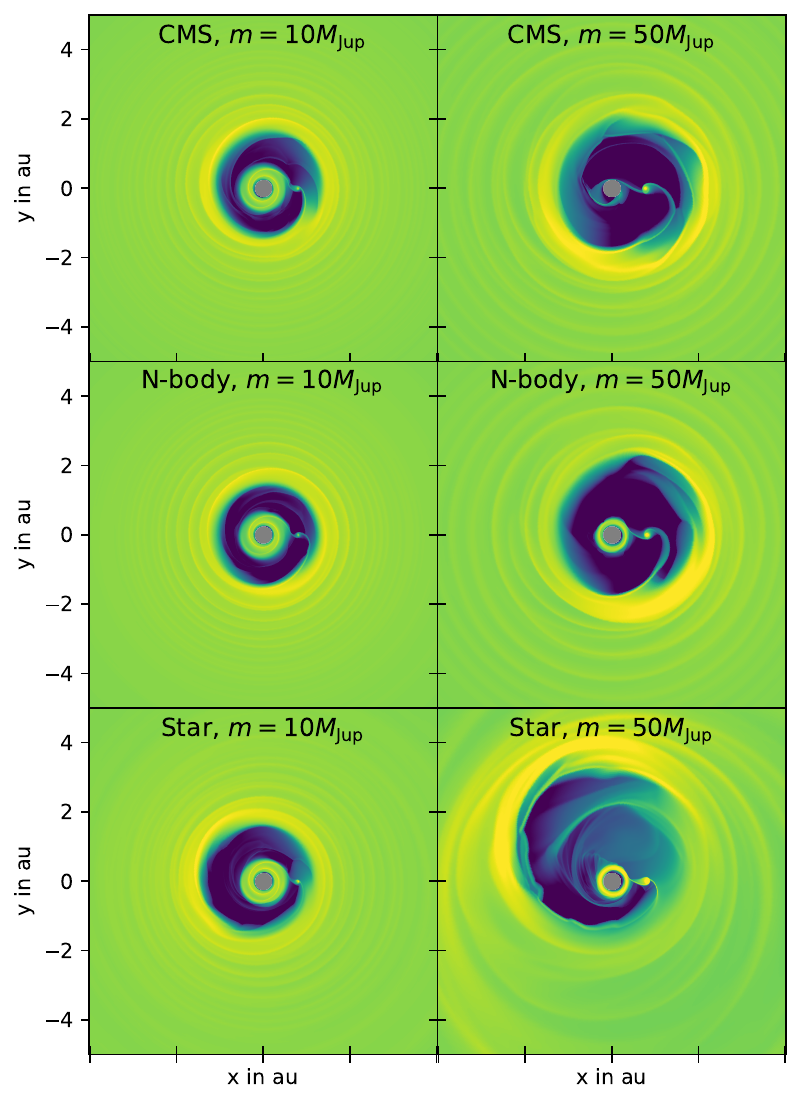}
		\caption{Snapshot of the logarithm of the surface density of a massless disk
		at the end of the initialization period ($t = 200~P_\mathrm{orb}$).
		"CMS" indicates that the simulation is in the center-of-mass system,
		"star" indicates that the simulation is centered on the star and the 
		disk has been initialized around the star. 
		"N-body" indicates that the simulation is centered on the star,
		but the disk has been initialized around the center of mass of the N-body system.
		All plots use the same color scale.
		\label{fig:init_instability}}
	\end{center}
\end{figure}

The way the disk is initialized can affect the long-term evolution of the disk.
In our previous runs, the simulations centered on the star 
were also initialized around the star so that the surface density
was a function of the distance of the cell from the star, and
the initial gas velocity was computed with the mass of the star.
This caused the disk to become eccentric for 
high companion masses ($m \gtrapprox 10 M_\mathrm{Jup}$)
as the companion ramped up in mass. An example of this is shown in
the third row of \figref{fig:init_instability}.
For $m = 10 M_\mathrm{Jup}$ (left side) these effects start to become 
noticeable, while for $m = 50 M_\mathrm{Jup}$ (right side) the entire disk
and the gap around the companion become eccentric.

This does not happen if the disk is initialized using the distance to the center of mass 
and the total mass of the N-body system to compute the velocities,
as demonstrated for simulations centered on the star (second row) and 
centered on the center of mass (first row).

For even more massive stellar-mass companions, it is no longer possible 
to ramp up the mass of the companion, because the missing gravity from the 
not yet fully ramped up companion would cause the disk to deform 
before the companion reaches its final mass. Without ramping, longer
initialization times are required because strong shocks occur 
during the clearing of the gap that depend on the artificial viscosity.
These initialization artifacts then resolve on viscous time scales.

The advantage of centering the disk on the primary instead of the center of 
mass is also demonstrated in \figref{fig:init_instability}. 
At high masses (top right panel), the primary orbits inside 
the inner radius, causing its stellar disk to quickly accrete
through the inner boundary. This does not happen when the simulation is 
centered on the primary (middle right panel), in which case the disk 
is optimally resolved by the grid, which is the main motivation for this study.

After an initialization phase of $200$ companion orbits, 
we enabled the interaction between the disk and the companion and continued the simulations, meaning 
that the companion migrates under the gravitational influence of the disk, while the 
gravity of the gas on the star was added to the indirect term of the simulation.
With feedback enabled, the forces exerted on the binary system can move its center of mass, 
so we shifted the N-body system to the new center of mass of the binary
after each time step.
This was not necessary during the initialization phase because the forces are turned off and the center of mass of the binary stays put.

\begin{figure}
	\begin{center}
		\includegraphics[width=\linewidth]{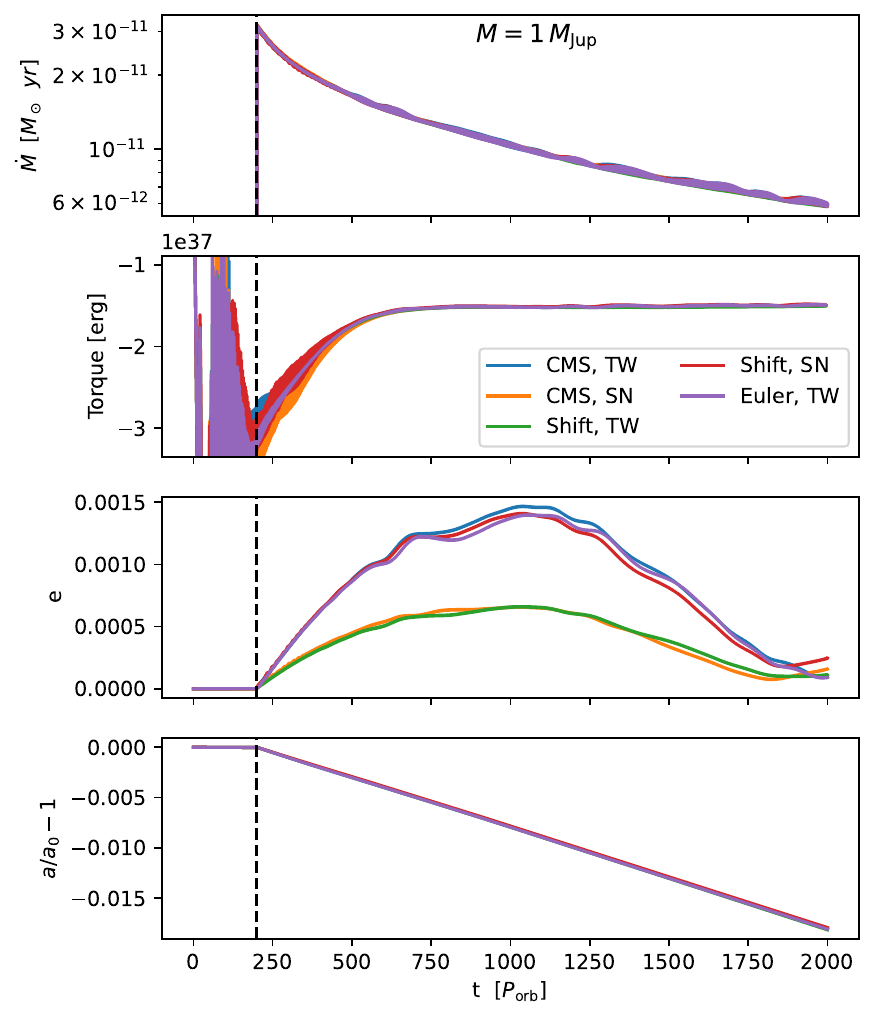}
		\caption{\label{fig:planet1} Various quantities measured for a 
		$m = 1~M_\mathrm{Jup}$ companion. Plotted are the mass accretion rate by the companion, 
		the gas torques exerted by the disk on the companion, the eccentricity,
		and the normalized semi-major axis of the companion.}
	\end{center}
\end{figure}

We then monitored the mass accretion rate on the companion, the torque exerted by the gas on the companion,
the eccentricity and the semi-major axis of the companion. The results for a $m = 1~M_\mathrm{Jup}$ companion
are shown in \figref{fig:planet1}. We find no significant differences
between the different artificial viscosities, indirect term protocols, or coordinate centers.
The gas torques oscillate for a longer time in the SN artificial viscosity simulations,
but they settle to the same value as in the TW artificial viscosity simulations and from then on
evolve the same.
Although the eccentricity variations display factor of 2
difference between the runs, the overall level of planetary eccentricity is too small ($\sim 10^{-3}$)
to be considered a significant effect, and the differences do not appear to
be related to the artificial viscosity or the indirect term protocol.

\begin{figure}
	\begin{center}
		\includegraphics[width=\linewidth]{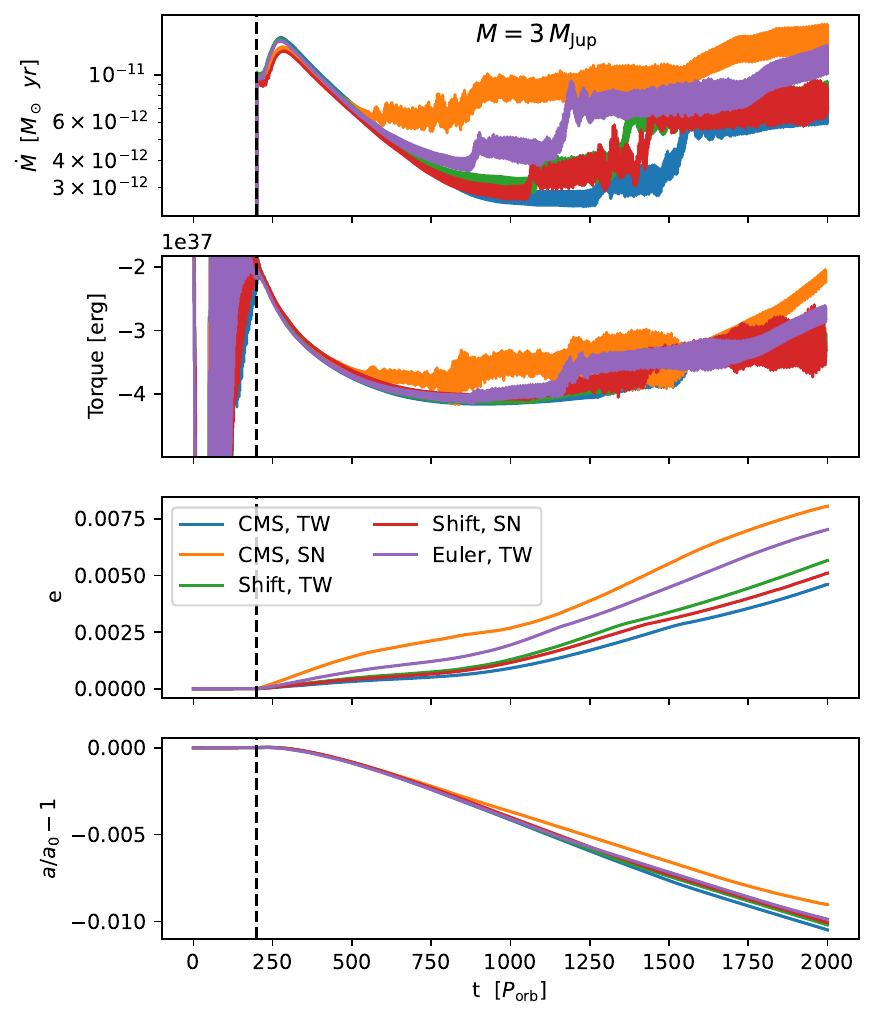}
		\caption{\label{fig:planet3} Same figure as \figref{fig:planet1} but 
		for a $m = 3~M_\mathrm{Jup}$ companion.}
	\end{center}
\end{figure}

For higher masses of $m = 3~M_\mathrm{Jup}$ and $m = 10~M_\mathrm{Jup}$,
we find small differences within the Hill sphere of the companion,
which we can directly attribute to the different
artificial viscosity prescriptions (cf. \figref{fig:massloss_sec_3}).
Globally, we find that the time at which the gap becomes eccentric 
is different for each setup. 
When the gap becomes eccentric, the density inside the gap is higher, leading
to higher accretion (see first panel in \figref{fig:planet3}) 
and the gas torque changes (see second panel). 
The differences in gas torque then lead to different eccentricities and 
migration rates of the companion.
Most of the differences between the setups can be explained by the different
time at which the gap becomes eccentric. We find found no trend in how the artificial viscosity
or the indirect term protocol affects the timing of this transition.
While the simulations are in the same state, they evolve similarly.

\begin{figure}
	\begin{center}
		\includegraphics[width=\linewidth]{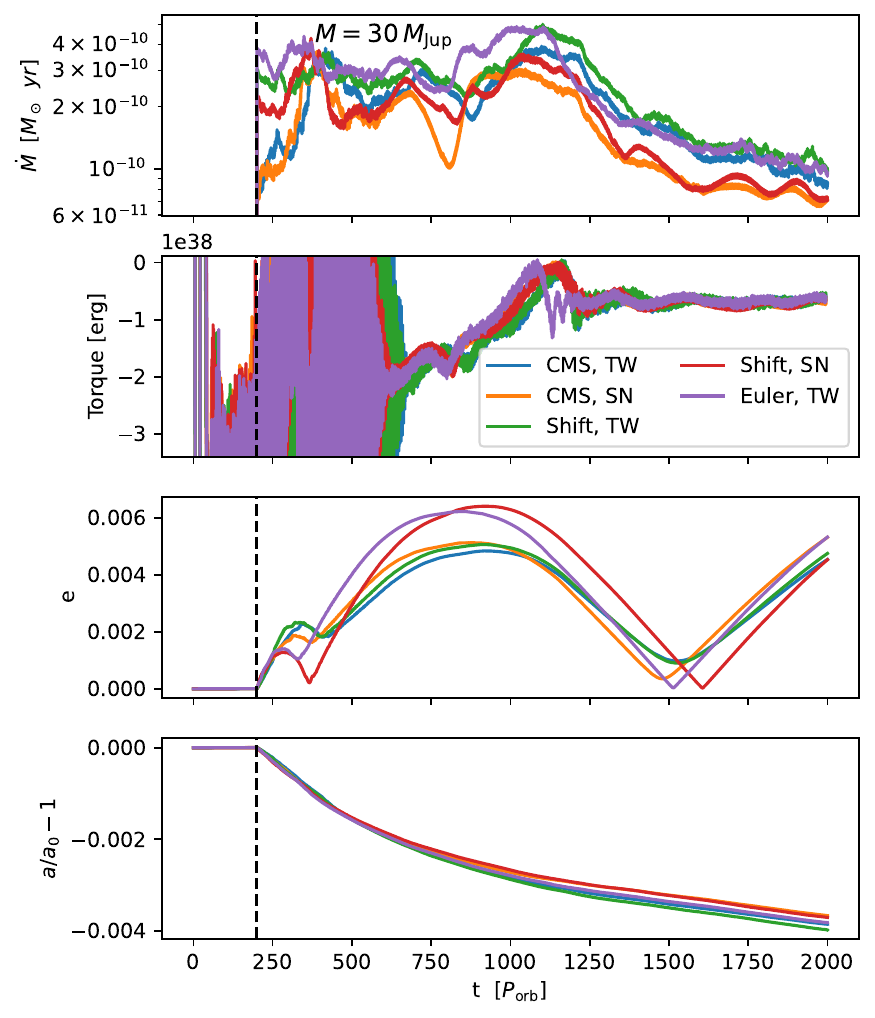}
		\caption{\label{fig:planet30} Same figure as \figref{fig:planet1} but 
		for a $m = 30~M_\mathrm{Jup}$ companion.}
	\end{center}
\end{figure}

For our high-mass companions ($m = 30~M_\mathrm{Jup}$ and $m = 50~M_\mathrm{Jup}$),
we find significant differences in mass accretion between the artificial viscosities,
as shown in \figref{fig:planet30}. 
The artificial viscosities eject mass from the Hill sphere of the companion 
(cf. \figref{fig:massloss_sec_50_setups}) resulting in lower mass accretion rates.
As depicted in the first panel of \figref{fig:planet30}, we find $\approx 50\%$
higher accretion rates and $\approx 50\%$ more mass inside the Hill sphere for the TW artificial viscosity.
This difference increases to almost $100\%$ for the $m = 50~M_\mathrm{Jup}$ companion.

Because we have ignored the gravitational effect of the gas inside its Hill sphere
on the companion by \eqref{eq:smooth}, the differences in gas torque 
are an indication of the effects of the different setups far away from the companion.
Again, we find that there are differences in gas torque and companion eccentricity
between the setups, but without a clear trend. This means that there is 
no strict increase or decrease in gas torque or eccentricity when changing
the artificial viscosity or the indirect term prescription.

However, in most of our simulations, we find that the companion migrates inwards
faster for the TW artificial viscosity. For example, in the last panel of \figref{fig:planet30}, 
we find $7\%$ faster inward migration for the TW artificial viscosity
than the SN artificial viscosity in the primary frame and $5\%$ faster in the center-of-mass frame.
We analyzed what causes this difference and found that in the 
center-of-mass frame, the gas torque on the planet is stronger for the TW artificial viscosity;
and in the primary frame, the indirect term due to the gas feedback into the central star,
which results in a positive torque on the companion, is weaker for the TW artificial viscosity.
At the end of the simulations ($t = 1800~T_\mathrm{orb}$ to $2000~T_\mathrm{orb}$),
the migration rates are similar for all setups.
It remains to be determined whether the faster inward migration observed with the
TW artificial viscosity represents a consistent trend or a coincidental effect.

\section{Summary and conclusions}
\label{sec:summary}
We have presented two modifications to the \textsc{Fargo} code \citep{masset2000fargo} that
allow us to simulate a circumbinary disk in the frame of one of the stars
which to our knowledge was not possible before.
One of the proposed modifications is the use of the tensor artificial
viscosity by \citet{tscharnuter1979artificial_viscosity} and the other is
a new method for calculating the indirect term.
We have also tested their relevance to simulations of planet-disk interactions.

First, we studied the indirect term prescription, that is,
the consideration of fictitious forces due to a non-inertial frame of reference.
In \secref{sec:nbody}, we noted that the standard method of applying the
indirect term is comparable to a forward Euler step and produces results
comparable to that of an Euler integrator. 
It fails to keep the central object at the coordinate center and similarly fails 
to keep other objects in the frame of the central object. 
We proposed to precompute the effective acceleration experienced by the central
object over the entire time step and use it as the indirect term.
This method is comparable to
advancing the whole system in time and then subtracting the velocity of the central
object at the end of the time step. This shift-based indirect term will by design keep
the velocity of the central object at the end of the time step at 0.
We then showed that at large radii, $r > 2~a_\mathrm{bin}$, it is significantly better 
at modeling the effects on a non-inertial frame of reference than the default method.

The proposed shift-based indirect term protocol should
also be applicable to other grid codes used to simulate the interaction
of disks with stars and planets.
For example, the PLUTO code has a hydrodynamics solver that uses a higher-order
time-stepping scheme, but in the versions known to the authors,
the synchronization between the hydrodynamics and N-body solvers is performed
once per integration time step and the indirect term is applied as a single Euler step.
It can be expected that synchronization at each substep would improve the accuracy
of the evolution of the combined N-body and hydrodynamics system. How this plays out in practice
remains to be seen. 

In \secref{sec:art_vis} we listed literature \citep{tscharnuter1979artificial_viscosity,
stone1992zeus2d,bodenheimer2006numerical}, which already noted that the 
artificial viscosity by \citet{neumann1950shocks}, which is the default
artificial viscosity in the \textsc{Fargo} \citep{masset2000fargo}
and \textsc{Fargo3D} codes \citep{benitez2016fargo3d}, is not suitable for 
curvilinear coordinates and can cause artificial pressure for 
smooth gas flows even in the absence of shocks. They also stated that the 
tensor artificial viscosity by \citetalias{tscharnuter1979artificial_viscosity}
should be preferred because it takes into account the geometry of the grid and is independent
of the frame of reference. We confirmed these claims and showed
that the artificial viscosity by \citetalias{neumann1950shocks}
inherently produces an artificial pressure in a smooth Keplerian disk,
which arises from the velocity gradients in the non-inertial frame of reference and scales with the number of azimuthal cells as $1/N_\varphi^2$.

In \secref{sec:binary} we simulated a circumbinary disk in the center of the lower-mass
companion of a binary system. In this setup, the indirect forces reach magnitudes
several times stronger than the direct gravitational forces of the N-body system, and
the standard indirect term prescription causes the disk to disperse. Our shift-based indirect term prescription keeps the disk 
stable, even at low resolutions, and allows the simulations to converge to a reference simulation
in the center-of-mass frame with increasing spatial and temporal resolution.

At low resolutions, the artificial pressure generated by the \citetalias{neumann1950shocks}
artificial viscosity caused by the velocity gradients in the non-inertial frame of reference,
was strong enough to prevent the disk from precessing, leading to significant eccentricity growth in the disk. 
At high resolutions ($> 8$ cells per scale height)
the simulation starts to converge for the different artificial viscosities.

We then tested the relevance of our changes to planet-disk interaction simulations in
\secref{sec:planet}. 
At low companion masses of $m \leq 1 M_\mathrm{Jup}$, 
we find no differences between the artificial viscosities or the indirect term protocols.
The default methods in the original \textsc{Fargo} code \citep{masset2000fargo}
are sufficient for these simulations.

For companion masses $m \geq 3 M_\mathrm{Jup}$ it becomes relevant to initialize the disk
in the center-of-mass frame instead of the center of the star. Otherwise, the shift 
of the center of mass caused by the ramping of the companion mass can cause the disk to become
eccentric. In addition, the artificial viscosity by \citetalias{neumann1950shocks} 
produces artificial pressure inside the Hill sphere of the companion (see also Appendix~\ref{app:companion_disk}).

For even higher masses, $m \geq 30 M_\mathrm{Jup}$, it does become necessary to initialize
the disk in the center-of-mass frame or the disk will become unstable as the companion mass
ramps up.
In addition, the artificial viscosity \citetalias{neumann1950shocks}
ejects significant amounts of mass from the Hill sphere of the companion,
resulting in lower mass accretion rates.
We also found differences in the gas feedback from the companion 
depending on the choice of artificial viscosity and indirect term protocol,
but these differences are small. 

In summary, we find that the methods used in the \textsc{Fargo} code \citep{masset2000fargo}
are sufficient for simulating planet-disk interactions in the frame of a central star,
but issues arise for more massive companion objects.
When the companion mass approaches the brown dwarf mass, 
the frame in which the disk is initialized has to be adjusted, and the 
artificial viscosity by \citet{neumann1950shocks} will cause artificial 
mass ejection from the Hill sphere of the companion. 
Initializing the disk in the center-of-mass frame would also be relevant
for any other grid code centered on the star.

For stellar mass companions, the artificial acceleration 
caused by the \citet{neumann1950shocks} artificial viscosity due to the indirect term
can cause eccentricity growth throughout the disk. 
Problems related to artificial viscosity can be mitigated by using
the tensor artificial viscosity by \citet{tscharnuter1979artificial_viscosity}.
In addition, for stellar mass companions, the indirect term becomes several times
stronger than the direct gravitational forces of the N-body system inside the
circumbinary disk. In such cases, more accurate indirect-term prescriptions
should be used. We believe that the shift-based indirect term protocol
presented in this work is a good choice when the indirect term is applied in a single step
per time step.

\begin{acknowledgements}
	TR acknowledges funding from the Deutsche Forschungsgemeinschaft (DFG) research group FOR 2634 ''Planet Formation Witnesses and Probes: Transition Disks''
	under grants KL 650/29-2.
	The authors acknowledge support by the High Performance and Cloud Computing Group at the Zentrum f\"ur Datenverarbeitung of the University of T\"ubingen, the state of Baden-W\"urttemberg through bwHPC and the German Research Foundation (DFG) through grant INST\,37/935-z1\,FUGG.\end{acknowledgements}

\bibliographystyle{aa}
\bibliography{references}

\begin{appendix}
\section{Companion disk}
\label{app:companion_disk}
In \secref{sec:planet} we measured the effects of the artificial viscosities 
far from the companion by excluding the torques inside the Hill sphere.
While the accretion rate depends on the amount of mass inside the Hill sphere, 
it is unclear how much it is affected by the different migration rates.

In this section, we isolate the effects 
of the artificial viscosities inside the Hill sphere of the companion.
To achieve this, we ran simulations with the same parameters as in \secref{sec:planet},
but kept the central star and companion on fixed orbits by disabling 
disk feedback. Additionally, we tried to remove the  
circumstellar disk by reducing the domain size
to $0.6\,\mathrm{au}$ to $1.5\,\mathrm{au}$
with a resolution of $N_r \times N_\varphi = 147 \times 1005$ (cell sizes remained the same).
Removing the circumstellar disk from the simulation isolates the gap region and, thus, 
we are able to avoid contributions to the mass of the Hill sphere that would otherwise stem from gas traversing the gap.
As before, we used strict outflow conditions at both boundaries.

To reach a steady state, we kept the mass inside the simulation domain constant
by rescaling the surface density of each cell after each time step
by a factor equal to the initial gas mass divided by the current gas mass.
This density rescaling function adds the outflowing mass preferentially to cells with high density.
Since gas feedback is not considered here, the surface density cancels out in
the momentum equations and the actual disk mass is not relevant.

\begin{figure}
	\begin{center}
		\includegraphics[width=\linewidth]{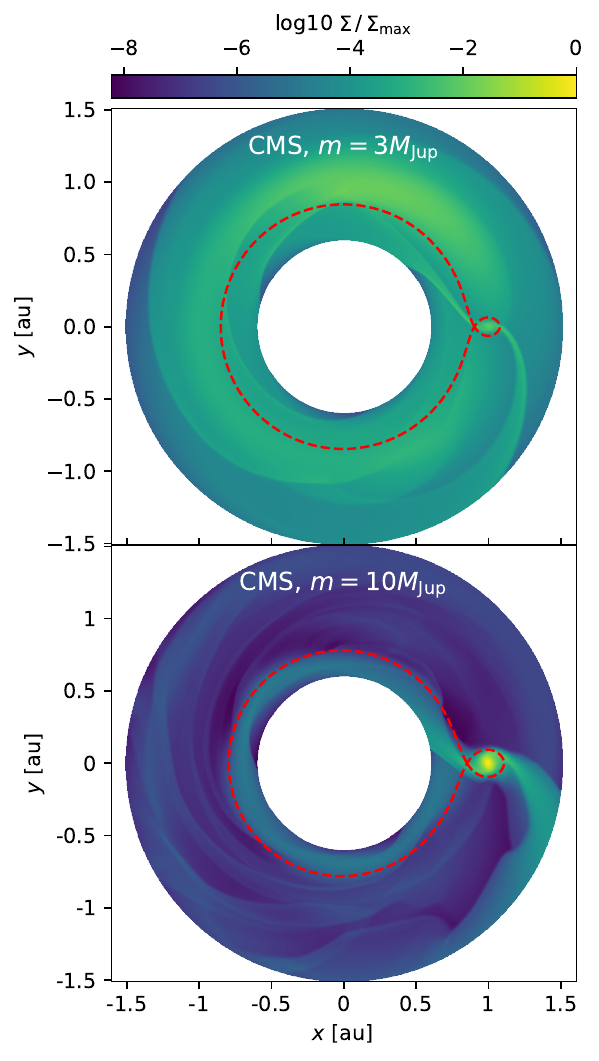}
		\caption{\label{fig:SigmaSec} Surface densities after 200 orbital periods
		for different companion masses. The dashed cyan line represents the 
		Roche lobe of the system.  Both simulations were run in the center-of-mass frame
		and used the TW artificial viscosity. Both plots use the same color scale.}
	\end{center}
\end{figure}

The surface density profiles after 200 orbital periods are shown in \figref{fig:SigmaSec}
for different companion masses, $m$.
In principle, the $3M_\mathrm{Jup}$ planet in the upper plot clears
its orbit, but because the mass is replenished by the density rescaling function,
the gap region appears to be filled with gas.
What the plot actually shows is the density ratio between the
gap region and the circumplanetary disk.
While this setup is not physical, it allows us to
compare the effects of the different numerical methods 
by measuring the amount of mass held inside the Hill sphere of the companion.

\begin{figure}
	\begin{center}
		\includegraphics[width=\linewidth]{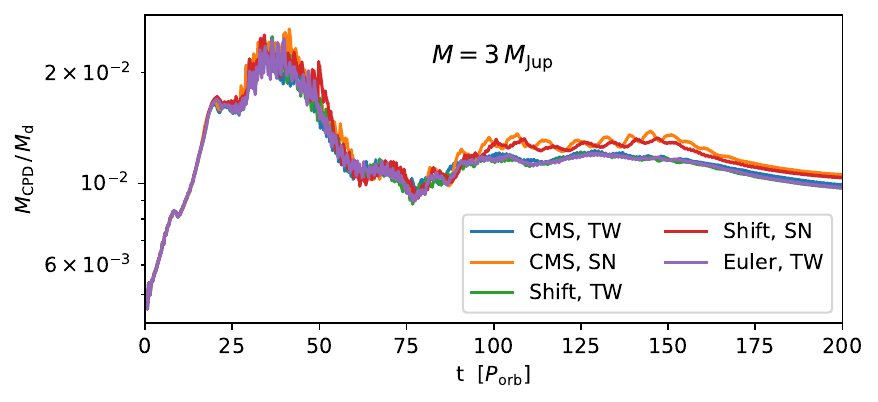}
		\caption{\label{fig:massloss_sec_3} Disk inside the Hill sphere of a $m = 3~M_\mathrm{Jup}$ companion
		divided by the total gas mass inside the simulation domain
		for different simulation setups.}
	\end{center}
\end{figure}

The normalized amount of mass inside the Hill sphere over time is shown in \figref{fig:massloss_sec_3}
for the $m = 3 M_\mathrm{Jup}$ companion. 
Neither the frame of reference nor the indirect term prescription affects the amount of 
mass around the companion, but for the artificial viscosities,
differences become visible after $100~P_\mathrm{orb}$.
At the end of the simulation, the Hill sphere contains 
$8\%$ more mass for the SN artificial viscosity.
We repeated this test for a $m = 1 M_\mathrm{Jup}$ companion, and found
a difference of only $0.3\%$ in mass inside the Hill sphere.
Similar to the long range effects probed in \secref{sec:planet}, we again 
find that the choice of artificial viscosity will affect the simulations at these companion masses.

For companion masses $m = 10 M_\mathrm{Jup}$ and above, the companion clears its orbit and
almost all of the gas mass accumulates inside the Hill sphere, see the lower plot in \figref{fig:SigmaSec}.
Since the mass inside the Hill sphere is the same for all setups, we measure the effect
of the artificial viscosities by how much mass is ejected from the companion disk through
the inner and outer boundaries of the domain.

\begin{figure}
	\begin{center}
		\includegraphics[width=\linewidth]{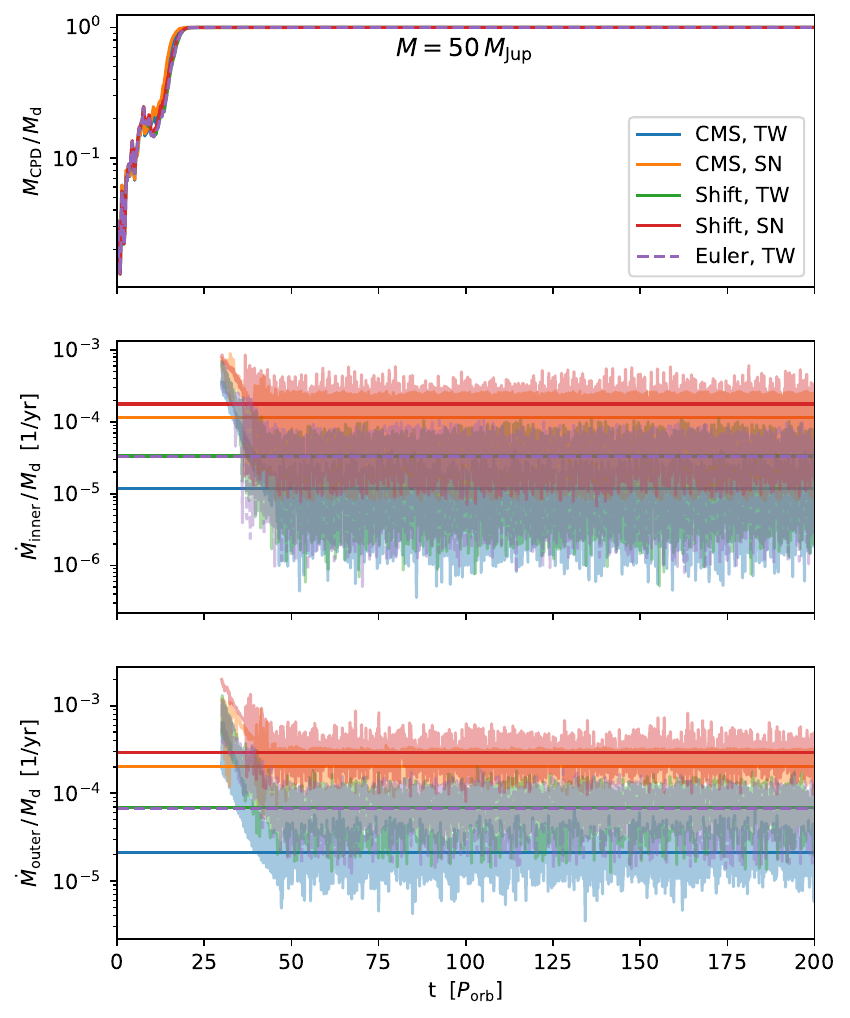}
		\caption{\label{fig:massloss_sec_50_setups} 
		Disk mass and mass loss rates over time for a $m = 50~M_\mathrm{Jup}$ companion
		for different setups.
		Top panel: mass inside the Hill sphere of the companion.
		Second panel: mass loss rate through the inner boundary.
		Bottom panel: mass loss rate through the outer boundary.
		All quantities are normalized to the total gas mass inside the domain.}
	\end{center}
\end{figure}

The two lower panels in \figref{fig:massloss_sec_50_setups} show the mass loss 
rates for the different setups with a $m = 50 M_\mathrm{Jup}$ companion. 
The effects of the indirect term protocol are still negligible.
The artificial viscosity again has the largest effect with the SN
artificial viscosity leading to 10 times more mass ejection 
than the TW artificial viscosity in the CMS 
and 5 times more in the primary frame simulations.
At this mass, the frame of reference also makes a difference 
with the primary frame simulations ejecting 3 and 1.5 times more 
mass than the CMS simulations for the TW and SN artificial viscosities, respectively.

Without the mass replenishment function, the companion disks would 
lose mass, leaving less mass for accretion and explaining the 
different mass accretion rates found in \figref{fig:planet30}.

For the $m = 10 M_\mathrm{Jup}$ the mass loss is higher for the TW artificial viscosity 
by a factor of $12\%$ and $100\%$ for the inner and outer boundary respectively, 
which is similar to the higher disk masses measured 
for the $m=3 M_\mathrm{Jup}$ companion in \figref{fig:massloss_sec_3}.
This implies that between 10 and 30 Jupiter masses, the SN artificial viscosity
becomes more active and changes from lower to higher mass ejection
than the TW artificial viscosity.


%
%

\end{appendix}

\end{document}